\documentclass[conference]{IEEEtran}
\IEEEoverridecommandlockouts
\usepackage{cite}
\usepackage{amsmath,amssymb,amsfonts}
\usepackage{algorithmic}
\usepackage{graphicx}
\usepackage{textcomp}
\usepackage{xcolor}

\makeatletter
\def\BibTeX{{\rm B\kern-.05em{\sc i\kern-.025em b}\kern-.08em
    T\kern-.1667em\lower.7ex\hbox{E}\kern-.125emX}}
\begin{document}

\title{Digital Twins for Internet of Battlespace Things (IoBT) Coalitions
\\

\thanks{Gemini AI (Google) was used for editing and grammar enhancement.}
}

\author{
	\centering
	\begin{tabular}{c}
    
    \begin{tabular}{ccc}
        Athanasios Gkelias & Patrick J. Baker & Kin K. Leung \\
        \textit{EEE Department} & \textit{Rapid Capabilities Office} & \textit{EEE Department} \\
        \textit{Imperial College London} & \textit{Royal Air Force} & \textit{Imperial College London} \\
        London, UK & Farnborough, UK & London, UK \\
        a.gkelias@imperial.ac.uk & pbaker@dstl.gov.uk & kin.leung@imperial.ac.uk \\
	\end{tabular}\\[1.5cm]
    \begin{tabular}{cc}
        Olwen Worthington & Christopher R. Melville \\
        \textit{Cyber and Information Systems} & \textit{Rapid Capabilities Office}\\
        \textit{Defence Science and Technology Laboratory} & \textit{Royal Air Force} \\
        Porton Down, UK & Farnborough, UK \\
        olworthington@dstl.gov.uk & christopher.melville489@mod.gov.uk
    \end{tabular}
   	\end{tabular}

}

\maketitle

\begin{abstract}

This paper presents a new framework for integrating Digital Twins (DTs) within Internet of battlespace Things (IoBT) coalitions. We introduce a novel three-tier architecture that enables efficient coordination and management of DT models across coalition partners while addressing key challenges in interoperability, security, and resource allocation. The architecture comprises specialized controllers at each tier: Digital Twin Coalition Partner (DTCP) controllers managing individual coalition partners' DT resources, a central Digital Twin Coalition(DTC) controller orchestrating cross-partner coordination, and Digital Twin Coalition Mission (DTCP) controllers handling mission-specific DT interactions. We propose a hybrid approach for DT model placement across edge devices, tactical nodes, and cloud infrastructure, optimizing performance while maintaining security and accessibility. The architecture leverages software-defined networking principles for dynamic resource allocation and slice management, enabling efficient sharing of computational and network resources between DT operations and primary IoBT functions. Our proposed framework aims to provide a robust foundation for deploying and managing Digital Twins in coalition warfare, enhancing situational awareness, decision-making capabilities, and operational effectiveness while ensuring secure and interoperable operations across diverse coalition partners.

\end{abstract}


\section{Introduction}

The Internet of Battlespace Things (IoBT) represents a great advancement in military technology, integrating diverse battlespace assets into a unified, intelligent network that enhances military capabilities and operational effectiveness. This interconnected ecosystem transforms traditional warfare by enabling seamless communication, coordination, and decision-making across all battlespace domains. Smart devices and sensors form the foundation of IoBT, incorporating ground and air sensors, wearable technology, and multimodal sensing capabilities that span audio, visual, infra-red, radar, and biological domains, creating a comprehensive sensing mesh that provides unprecedented situational awareness across the battlespace. 

In addition, Digital Twin (DT) technology has emerged as a key enabler in modern battlespace operations, providing sophisticated virtual representations of battlespace assets, systems, and operations. These virtual replicas enable real-time monitoring, simulation, and optimization, allowing military commanders to test scenarios, optimize strategies, and predict potential outcomes before actual deployment. The integration of DTs with IoBT creates a powerful synergy that enhances military capabilities across all operational domains.

On the other hand, coalition operations have become increasingly vital in modern warfare due to the growing reliance on shared technological capabilities. NATO's operations in Afghanistan, for example, underscored the critical need for interoperable systems and shared situational awareness among coalition partners. Similarly, multinational exercises like RIMPAC (Rim of the Pacific Exercise) highlight the importance of interoperability in maritime contexts. These real-world examples demonstrate the need for sophisticated technological solutions that enable effective collaboration while maintaining security and operational efficiency. Shared IoBT and DT technologies can significantly enhance the efficiency, effectiveness, and collective capabilities of such coalitions. 

Recent research has focused on the integration of DTs in military operations and coalition environments. NATO's MSG-205 Research Task Group has been investigating DT interoperability challenges across allied forces, working to establish common standards and frameworks for DT implementation in multinational operations. The group's work emphasizes the need for standardized approaches to data sharing, security protocols, and model interoperability across coalition partners~\cite{Skinner2023}. Poltronieri et al.\cite{Poltronieri2018} proposed a device-agnostic architecture for IoBT device monitoring and control. By identifying devices based on capabilities rather than type, this approach simplifies client-side control and enables platform-independent operations.\cite{Wigness2023} detailed the challenges posed by the scale, heterogeneity, information sharing needs of coalition environments, dynamic behaviour, and sophisticated adversaries, while also exploring emerging directions for scalable, secure, and performant IoBT.
\cite{Strayer2023} presented two architectures and their corresponding trade-offs for content-centric military IoT, optimizing information dissemination across tactical data links for disconnected, intermittent, and limited (DIL) connectivity.
\cite{Liu2023} proposed a digital thread and twin-based collaborative manufacturing framework for real-time monitoring and adaptive task adjustment in industrial IoT, using a heterogeneous information network for unified lifecycle data and digital threads for product state analysis. \cite{Hui2022} proposed a DT-enabled collaborative and distributed autonomous driving scheme that leverages auction game-based collaborative driving and coalition game-based distributed driving mechanisms to optimize group formation and minimize autonomous driving costs while reducing physical network communication overhead. \cite{Attaran2023} addressed the challenges of DT implementation in resource-constrained environments by proposing architectures that balance model accuracy, computational limitations, security, and operational effectiveness. \cite{Reim2022} proposed a contingency framework for managing digital platform collaboration in manufacturing, focusing on Digital Twin platforms, by analysing challenges and management approaches through an in-depth case study, providing practical guidance for implementing successful platform-based collaborations.

This paper presents a novel framework for integrating DTs within IoBT coalitions. The primary contributions include the introduction of a novel three-tier architecture that enables efficient coordination and management of DT models across coalition partners. This architecture addresses key challenges in interoperability, security, and resource allocation by incorporating specialized controllers at each tier: Digital Twin Coalition Partner (DTCP) controllers for managing individual partners' DT resources, a central Digital Twin Coalition (DTC) controller for cross-partner coordination, and Digital Twin Coalition Mission (DTCM) controllers for mission-specific DT interactions. Additionally, the paper discusses a hybrid approach for DT model placement across edge devices, tactical nodes, and cloud infrastructure, optimizing performance while maintaining security and accessibility. Finally, the framework leverages software-defined networking principles for dynamic resource allocation and slice management, enhancing situational awareness, decision-making capabilities, and operational effectiveness in coalition warfare.

The remainder of this paper is organized as follows: Section~\ref{sec:IoBT} provides an overview of IoBT and its key technologies. Section~\ref{sec:DT} examines DTs and their role in IoBT environments, including benefits and implementation challenges. Section~\ref{sec:coal} explores IoBT coalitions and how DTs can enhance coalition operations. Section~\ref{sec:arch} presents our proposed architecture for DT integration in IoBT coalitions, detailing model placement strategies, controller frameworks, and software-defined resource management. Finally, Section~\ref{sec:disc} discusses future research directions, while Section~\ref{sec:con} concludes the paper.

\section{Internet of Battlespace Things (IoBT)}\label{sec:IoBT}

By integrating diverse battlefield assets into a unified, intelligent network, IoBT offers a revolutionary advancement in military technology.  This interconnected ecosystem enables seamless communication, coordination, and decision-making across all domains, significantly enhancing military capabilities and operational effectiveness. Smart devices and sensors form the foundation of IoBT, incorporating ground and air sensors, wearable technology, and multimodal sensing capabilities that span audio, visual, infra-red, radar, and biological domains. These sensors create a comprehensive sensing mesh that provides unprecedented situational awareness across the battlespace. Working in concert with these sensors, robotics and autonomous vehicles, including drones, unmanned vehicles, and swarm intelligence systems, extend the reach and capabilities of military forces while minimizing human exposure to dangerous situations. 

Artificial Intelligence (AI) serves as the cognitive backbone of IoBT, enabling autonomous decision-making in real-time, predictive analysis, and self-optimization capabilities. AI systems also provide self-healing and resilience features, ensuring continued operation even under adverse conditions. This intelligence layer is supported by advanced connectivity solutions, particularly 5G and beyond, which provide the high-speed, low-latency communications essential for real-time operations. Software Defined Network (SDN) slicing and customization capabilities ensure that critical communications receive appropriate prioritization and resources. Edge and cloud computing architectures work in tandem to process and analyse battlespace data. Edge computing enables real-time processing and reduced bandwidth dependency at the tactical edge, while cloud computing provides the scalability and advanced analytics capabilities needed for complex operations. This hybrid approach ensures both immediate tactical response and strategic depth in data analysis. The integration of Augmented Reality (AR) and Virtual Reality (VR) technologies enhances situational awareness and training capabilities, while cybersecurity and blockchain technologies ensure secure data sharing across the battlespace network. Advanced computing paradigms, including neuromorphic and quantum computing, promise to push the boundaries of computational capabilities in battlespace environments. All these technologies, working in concert, create a sophisticated and resilient battlespace network that enhances military capabilities across all domains. Their synergistic integration enables more effective, efficient, and safer military operations while providing commanders with unprecedented levels of control and situational awareness.

On the other hand IoBT faces several critical challenges that must be addressed to ensure its effective deployment and operation in military environments. At the forefront of these challenges is cybersecurity, as the interconnected nature of battlespace systems makes them potentially vulnerable to sophisticated hacking attempts and cyberattacks, which could compromise mission-critical operations. This security concern is closely intertwined with the challenge of interoperability, as the diverse array of systems and platforms from different manufacturers must work together seamlessly while maintaining robust security protocols. The harsh realities of battlespace environments further compound these challenges through power and battery life limitations, requiring innovative solutions to extend the operational duration of devices in adverse conditions. Additionally, the massive amount of data generated by IoBT devices demands sophisticated network infrastructure capable of handling high data flows with minimal latency, while simultaneously addressing power constraints for sustained operations. These interconnected challenges highlight the complexity of implementing IoBT systems and underscore the need for integrated solutions that address security, interoperability, power management, and network performance simultaneously.

\section{Digital Twins for IoBT}\label{sec:DT}

Digital Twins (DTs) — virtual replicas of physical systems, assets, processes, capabilities, and even personnel — enable real-time monitoring, simulation, and optimization. In military applications, this translates to sophisticated virtual representations of battlespace assets, systems, and operations, continuously updated with real-time data to reflect current conditions and enable predictive capabilities. The integration of DTs with the IoBT unlocks a wide range of transformative benefits for military operations, impacting situational awareness, decision-making, resource allocation, and mission planning. For situational awareness, DTs provide real-time, multidimensional visualizations of the battlespace, giving commanders comprehensive insights. Predictive analytics, using both historical and real-time data, enables proactive identification of threats and opportunities. Optimized resource allocation is achieved by analysing real-time data and simulating various scenarios. DTs facilitate improved decision-making through advanced simulation and modelling, allowing planners to evaluate strategies virtually, while virtual mission rehearsal capabilities revolutionize mission planning and execution by identifying risks and optimizing tactics. Finally, DTs enable adaptive planning, ensuring mission flexibility in dynamic environments.

DTs impact the entire lifecycle of military technologies and operations. From accelerating innovation through rapid prototyping and continuous improvement of equipment and tactics, to enhancing operational efficiency through predictive maintenance and optimized logistics, DTs provide significant advantages. They ensure optimal performance, minimize downtime, and guarantee timely support for military operations. Furthermore, DTs significantly enhance warfighter safety through comprehensive risk assessment. By identifying potential hazards, commanders can implement proactive mitigation measures. DTs also accelerate emergency response by providing real-time situational information, enabling faster, more effective reactions to emerging threats or incidents. This comprehensive DT and IoBT integration creates a powerful synergy, enhancing military capabilities across all operational domains — from tactical decisions to strategic planning — while prioritizing warfighter safety and mission success.

However, implementing DTs within IoBT environments presents significant challenges, primarily related to resource constraints.
A key challenge is balancing DT model accuracy with complexity, i.e., more accurate models require significantly more real-time, high-frequency data collection and processing, which strains available resources. This is exacerbated by the distributed nature of IoBT, where the high bandwidth and low latency required for complex DTs may not always be available. Competition for computational and network resources between physical IoBT systems and their digital twins further compounds the issue. Operating in hostile environments introduces security vulnerabilities, especially with intermittent or compromised communications, making it difficult to maintain DT functionality. These challenges are particularly pronounced at the tactical edge, where limited processing power and network connectivity clash with the critical need for real-time DT updates and accurate modelling.

\section{DT for IoBT Coalitions}\label{sec:coal}

IoBT coalitions represent complex interconnected networks where multiple military forces share technological resources and capabilities to achieve common operational objectives. In a NATO coalition scenario, member nations might integrate their diverse IoBT assets - such as advanced sensor networks from the US, drone swarms from the UK, and autonomous ground vehicles from Germany - all coordinated through a unified command and control structure. These forces would share critical technologies including AI-driven decision support systems, secure communication networks, and DT models while maintaining appropriate security protocols and national sovereignty over sensitive capabilities. In contrast, a national coalition scenario might involve different branches of a single nation's military, such as the Army, Air Force, and Navy, combining their IoBT resources more comprehensively. For example, the Air Force's aerial surveillance platforms could seamlessly integrate with Army ground sensors and Navy maritime systems, sharing data through a common cloud infrastructure and leveraging shared AI analytics capabilities. This national coalition would benefit from easier integration due to common standards and protocols, enabling more comprehensive sharing of DT models, predictive maintenance systems, and tactical edge computing resources. In both scenarios, the coalition's effectiveness relies heavily on the sophisticated orchestration of shared IoBT resources through technologies like 5G communications, edge computing, and blockchain-secured data sharing, all while maintaining appropriate levels of access control and security.

\subsection{IoBT Coalition}

Creating an IoBT coalition where multiple defence partners share their IoBT resources for a common mission offers significant benefits over a scenario where each partner operates their IoBT independently. These benefits can be categorized into operational efficiency, enhanced situational awareness, resource optimization, and interoperability.

Sharing IoBT data allows for the integration of diverse sensor and information streams, providing a comprehensive, real-time situational awareness picture. This improves decision-making by reducing blind spots and increasing the accuracy of intelligence. Collaboration enables data fusion across different domains (e.g., air, land, sea, cyber, and space), allowing for insights that would not be possible with isolated IoBT systems. When properly coordinated, a coalition avoids duplication of sensor coverage and communication infrastructure, leading to more efficient use of available resources. Moreover, it allows the coalition to dynamically allocate resources (e.g., sensors, processing power, or communication bandwidth) based on mission priorities. This ensures critical areas or tasks receive adequate attention without overburdening any single partner. Data processing, storage, and communication loads can be distributed across coalition members (i.e., load balancing), enhancing the overall system's performance and resilience. 

In addition, each partner's IoBT may have unique capabilities (e.g., specialized sensors, AI algorithms, or communication systems). Combining these strengths creates a system that is more powerful and versatile than the sum of its parts. Sharing IoBT assets provides redundancy and robustness against failures, so if one partner's systems are compromised or fail, others can compensate to maintain mission continuity. Sharing IoBT infrastructure spreads costs and reduces the financial and operational burden on individual partners, particularly for smaller or resource-constrained entities.

However, the implementation of IoBT coalitions faces numerous complex challenges that span technical, operational, and organizational domains. At the core of these challenges lies the fundamental issue of system integration and interoperability. Coalition members typically operate diverse IoBT systems with varying hardware, software, and communication protocols, making the creation of a unified operational framework particularly challenging. This technical diversity is further complicated by the need to establish common standards for data exchange and communication protocols, a process that often involves sensitive political negotiations and time-consuming consensus-building among coalition partners. 

Security and trust management represent another critical challenge layer in IoBT coalitions. The varying levels of trust between coalition members directly impact how resources and data are shared and accessed. The challenge extends beyond simple access control to encompass sophisticated cybersecurity concerns, as a security breach in any partner's system could potentially compromise the entire coalition network. This interconnected nature of IoBT systems demands a coordinated approach to threat detection and mitigation, even as coalition partners maintain different security policies and tools. 

Moreover, resource allocation and data management present another significant operational challenges. Coalition members must dynamically share limited resources such as bandwidth, processing power, and sensor feeds while balancing mission objectives with fairness considerations. On the other hand, the integration of data from multiple sources, each with different formats, accuracy levels, and reliability, complicates the creation of a coherent operational picture. This challenge is amplified by the need to respect national security restrictions and privacy concerns while maintaining effective data sharing and operational coordination. Governance and scalability issues further complicate IoBT coalition operations. Clear protocols must be established for decision-making regarding shared resources, particularly in scenarios requiring real-time arbitration of competing priorities. 

Furthermore, the dynamic nature of military operations demands systems capable of rapidly integrating new coalition partners or additional IoBT resources without disrupting ongoing missions. This scalability requirement is particularly challenging when coalition members possess varying levels of technological sophistication or when temporary coalitions must be formed quickly in response to emerging threats. Accountability and performance evaluation present unique challenges in the coalition context. Tracking the contribution of individual coalition members' IoBT systems to mission outcomes, as well as identifying the source of errors or failures within the integrated system, becomes increasingly complex as systems become more tightly integrated. These challenges are particularly acute in scenarios requiring rapid adaptation to changing mission requirements or when dealing with temporary coalition arrangements that may lack established performance metrics or evaluation frameworks.

\begin{figure*}[htbp]
\centerline{\includegraphics[width=0.9\textwidth]{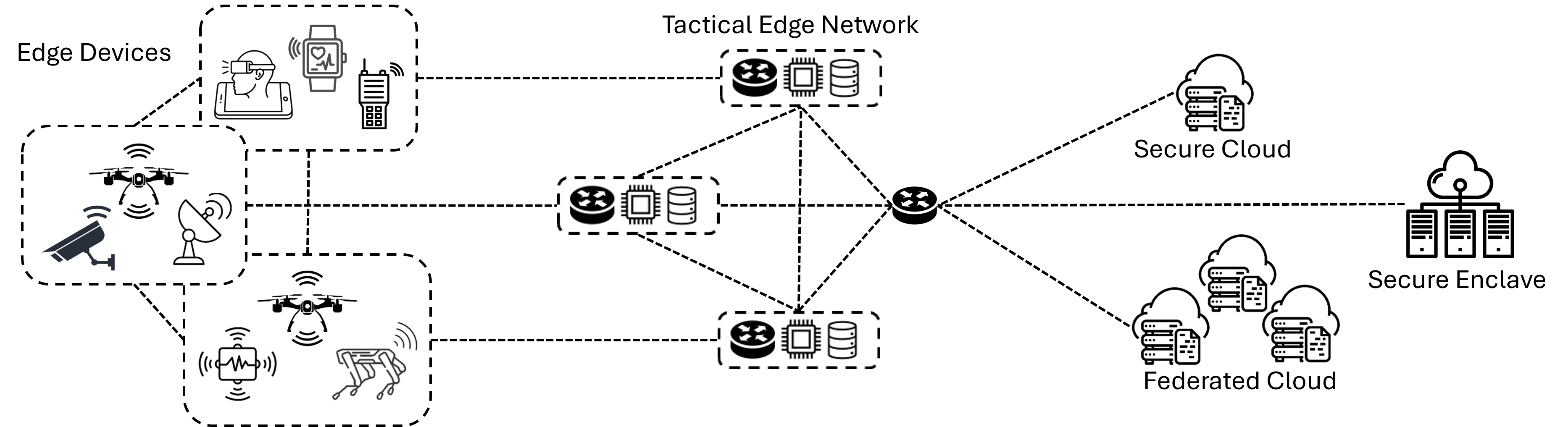}}
\caption{Three-tier IoBT network architecture illustrating the hierarchical deployment of battlespace systems across Edge Devices (including sensors, drones, and wearable technology), Tactical Edge Network (featuring distributed mesh nodes), and Cloud Infrastructure (comprising secure cloud, federated cloud, and secure enclave components).}
\label{figIoBT}
\end{figure*}

\subsection{DTs for IoBT coalitions}
DTs can serve as powerful tools for mitigating some of the complex challenges faced by IoBT coalitions. These virtual environments provide a safe space for pre-deployment testing and validation, allowing partners to model diverse IoBT systems, identify compatibility issues, and test integration strategies without impacting live networks. Within this secure sandbox, partners can simulate access control scenarios, test data-sharing protocols, and evaluate cybersecurity measures, while also facilitating consistent and reliable data processing and validation from diverse sources to ensure consistency and reliability in the coalition's operational picture.

Furthermore, critical resources such as bandwidth, processing power, and storage can be managed and utilized far more efficiently within IoBT coalitions through the use of DT simulations. These simulations allow partners to optimize resource allocation across their combined systems by testing dynamic resource allocation strategies and arbitration policies under various operational scenarios, ensuring efficient utilization during actual missions.

Moreover, DTs can offer practical solutions to the governance and scalability challenges faced by IoBT coalitions. By offering a risk-free environment to simulate decision-making processes and test governance policies, DTs enable partners to evaluate the impact of various operational scenarios on resource sharing and mission effectiveness, minimizing the potential for conflict during live operations. Importantly, DTs also provide a platform for testing the integration of new coalition members and simulating the impact of scaling operations, ensuring smooth adaptation to evolving mission requirements.

Most importantly, DTs enable continuous improvement and learning within IoBT coalitions. Through a shared platform for collaboration, testing, and problem-solving, partners can iteratively refine strategies and resolve operational challenges without impacting live systems. This platform also supports performance attribution and error tracking, enabling accurate assessment of contributions and efficient issue resolution within integrated systems.

However, DT coalitions in military environments face complex challenges spanning technical, operational, and organizational domains. The core challenge is interoperability across diverse partner systems, as each member typically employs different DT architectures, data models, and simulation frameworks, hindering seamless integration. This technical diversity is compounded by variations in computational capabilities and network conditions, potentially causing latency and time-drift issues that impact coordinated operations. Scaling DT coalitions becomes particularly difficult when integrating new partners or managing systems with varying levels of sophistication.

Security and trust management pose another critical set of challenges for DT coalitions. Diverse security protocols and access control policies among partners complicate secure data sharing while protecting sensitive operational information. The distributed nature of coalition DT systems further complicates verifying the accuracy and integrity of shared outputs without compromising proprietary information. This challenge is especially significant when decisions are based on combined simulations across multiple partners' systems.

Last but not least, careful consideration is required to address the unique governance and resource management challenges facing DT coalitions. Coordinating decision-making across distributed DT systems necessitates clear governance structures and policies, which can be difficult to establish given the autonomous nature of coalition partners. Resource arbitration, including the sharing of computational power and bandwidth, often leads to complex negotiations and potential operational delays. Ensuring sustainable and effective collaboration requires careful consideration of equitable cost-sharing arrangements for coalition-wide DT activities, as well as addressing data ownership and accountability.

\section{A DT Architecture for IoBT Coalitions}\label{sec:arch}
This section presents a logical architecture for enabling Digital Twin Coalitions (DTCs) within IoBT environments. The architecture's foundation rests on specialized DT-controllers that coordinate and manage various DT models contributed by different coalition partners while addressing interoperability challenges across model, data, and policy levels. Our approach considers three key aspects: First, we examine the optimal placement of DT models across the network, balancing critical factors such as security, performance, and accessibility. Second, we introduce a novel framework for IoBT DT coalitions that details the roles and interactions of different controllers. Finally, we demonstrate how software-defined resource slicing (SDS) can facilitate the deployment, integration, and management of DTs within IoBT coalitions, with particular emphasis on addressing scalability and interoperability challenges.

\subsection{DT model placement}

The optimal deployment of DT models in an IoBT coalition system requires a carefully orchestrated hybrid approach across three distinct layers [Fig.~\ref{figIoBT}], each supporting different types of DT models based on their capabilities and requirements.

In situations with limited or unavailable network communication, deploying lightweight DT models on edge devices (such as sensors, drones, and soldier equipment) becomes particularly important, enabling local data aggregation, preprocessing, and real-time decision-making. However, due to the limited processing power, memory, and battery life of these devices, careful resource management and optimization of the DT models are essential to prevent them from impacting the device's primary function.

Deploying DT models on tactical network backbone nodes (e.g., command posts, software-defined mesh networks, or mobile data centres) enables intermediate data processing and analytics, allowing for local optimizations and a more holistic view and coordination of nearby assets and operations. While these nodes offer greater resources than edge devices, they also have limitations. Their localized perspective, while broader than that of edge devices, can still lead to a fragmented view of the overall battlespace. Additionally, these nodes can become bottlenecks if compromised or overloaded, impacting both DT performance and other network functions.

Long-term data storage and historical analysis, including complex deep learning models, can be achieved by complex and resource intensive DT models placed in secure cloud environments owned by a single partner or federated cloud architectures where DT models are distributed across multiple partner clouds. Highly sensitive DT models should be deployed in high-security enclaves in order to enhance protection for critical information and operation. While cloud environments offer vast resources for complex DT models and long-term data storage, communication latency between the tactical edge and the cloud can be significant, hindering real-time applications.  

Therefore, a carefully orchestrated hybrid scheme is essential for effective DT integration within an IoBT coalition. This approach strategically distributes DT models across edge devices, tactical nodes, and cloud environments, ensuring seamless data sharing and interoperability between different DT instances and across all coalition partners. By leveraging the strengths of each network tier, this hybrid model can create a comprehensive, responsive, and secure DT ecosystem that enhances coalition operations across the entire battlespace.

\begin{figure*}[htbp]
\centerline{\includegraphics[width=0.9\textwidth]{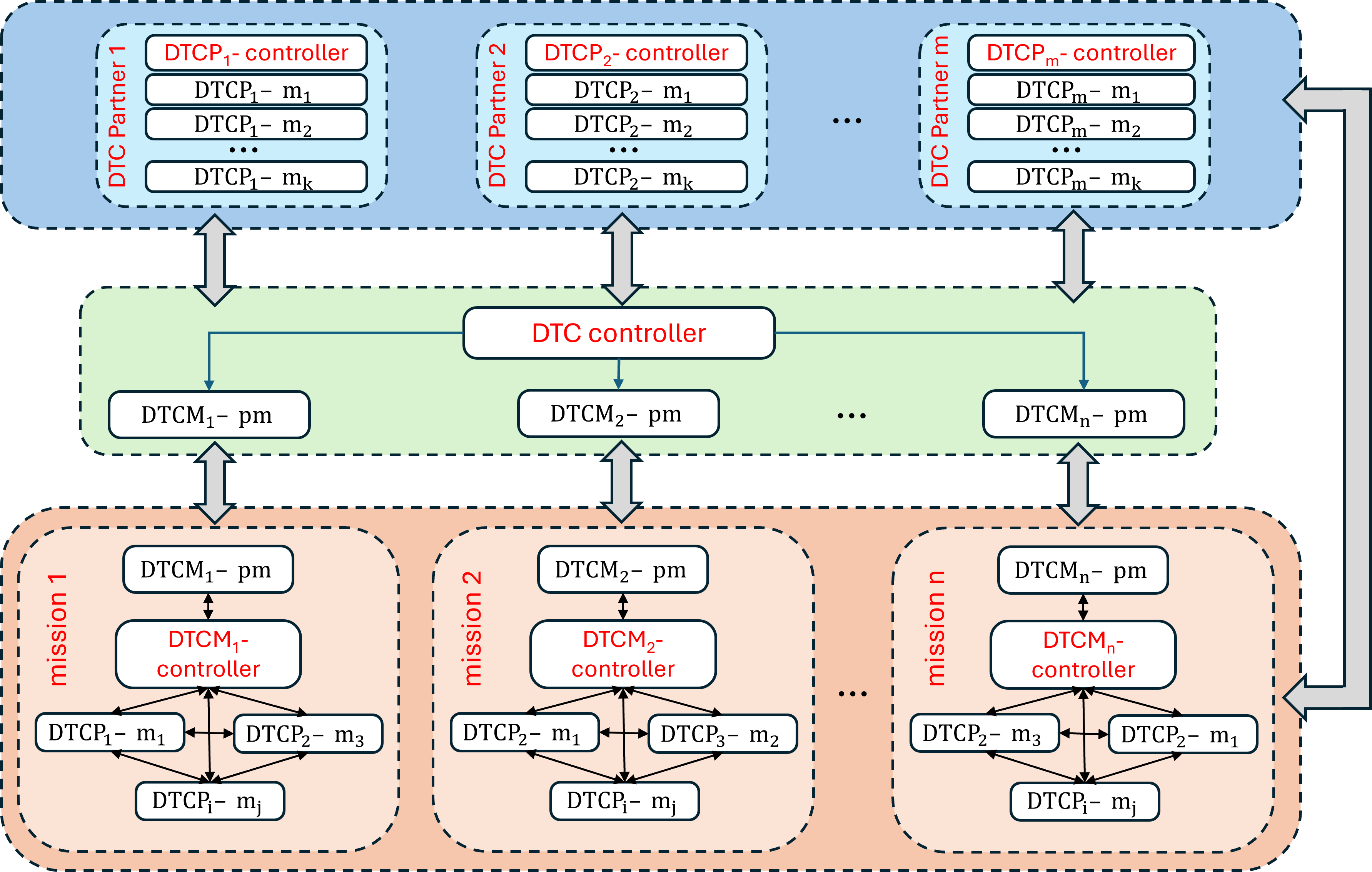}}
\caption{Digital Twin Coalition (DTC) Logical Architecture. The architecture consists of three hierarchical tier: (1) The Coalition Partners tier (blue) comprises multiple DTC Partners ($DTCP_1$ to $DTCP_m$), each with their own controller and multiple DT models ($m_1$ through $m_k$); (2) The coordination tier (green) features the central DTC controller coordinating across partners through Digital Twin Coalition Mission (DTCM) process models ($DTCM_1-pm$ through $DTCM_n-pm$); (3) The Mission Execution tier (orange) represents individual missions (1 through $n$), each managed by a DTCM controller that orchestrates interactions between selected DT models from different partners. Bidirectional arrows indicate communication flows between layers and components, enabling seamless coordination throughout the coalition structure}
\label{figArch}
\end{figure*}

\subsection{DT Architecture}

The proposed Digital Twin Coalition (DTC) architecture [Fig.~\ref{figArch}] presents a hierarchical framework designed to manage digital twins across IoBT coalition operations. This framework comprises three tiers, each with specialized controllers performing distinct yet interconnected functions, ensuring comprehensive and coordinated DT management.

\begin{itemize}
\item The coalition partners tier (blue section) comprises multiple Digital Twin Coalition Partners ($DTCP_1$ through $DTCP_m$), each managed by a dedicated DTCP-controller. These controllers maintain comprehensive registries of available DT models ($DTCP-m_1$ through $DTCP-m_k$) within each coalition partner's domain. They document crucial information about each DT model, including their capabilities, interfaces, and data requirements. This registry system enables efficient discovery and selection of appropriate DTs for specific coalition missions, ensuring that each partner's capabilities are properly catalogued and accessible for coalition operations. 

\item The DTC-controller, located at the coordination level (green section), serves as the primary orchestrator for the entire coalition. Connected to all DTCP-controllers, it coordinates DT model sharing and potential federation across all partners. Using inputs from corresponding Digital Twin Coalition Mission (DTCM) process models ($DTCM-pm$)—DT blueprints of mission objectives, operations, and processes — the DTC-controller determines optimal combinations and deployment of DT models for specific missions. 

A key function of the DTC-controller is ensuring interoperability. To address this, the DTC-controller may employ DT model transformation and adaptation techniques, orchestrate model composition and federation, enforce standardized interfaces with semantic mediation, and offer data mediation and transformation services to ensure consistent data quality, secure communication, and data synchronization for a unified view. Furthermore, it can enforce and resolve policy conflicts related to data sharing, access control, and security, translating and adapting policies and managing trust relationships. Finally, the DTC-controller can play a crucial role in pre-mission activities, facilitating mission rehearsal, resource optimization, and mission planning using selected coalition DT models.

\item The mission execution tier (orange section) comprises individual missions (1 through $n$), each managed by a dedicated Digital Twin Coalition Mission (DTCM) controller. These controllers manage the real-time interactions between different DT models during mission execution. They ensure seamless interoperability between diverse DT models by addressing challenges across models, data, and policies. They maintain communication with the DTC and related DTCP controllers when bandwidth permits, providing updates on DT model states and facilitating model updates and predictive maintenance.
\end{itemize}

This hierarchical control structure creates a robust framework that enables efficient discovery, federation, and coordination of DT models across coalition partners while maintaining operational flexibility and security. The architecture's emphasis on interoperability at the model, data, and policy levels ensures effective collaboration in complex coalition environments, ultimately enhancing decision-making and optimization capabilities in IoBT operations.

\subsection{Software Defined Slicing for DTCs}

Resource allocation for DTs in IoBT environments is challenging due to competition for limited computational and network resources (processing power, memory, storage, and bandwidth) between DT models and primary IoBT devices and services. This competition is especially acute at the resource-constrained edge layer, where devices must balance DT operations with their primary battlespace functions. In coalition environments, the challenge is compounded by multiple partners' DTs requiring access to shared resources and the need for frequent reallocation due to dynamic battlespace conditions. Software Defined Networking (SDN) offers a promising solution by providing flexible, programmable network control and resource management, enabling dynamic allocation based on real-time requirements and priorities. Through centralized control and network virtualization, SDN ensures critical battlespace operations maintain necessary resources while allowing DT models to effectively utilize available capacity.

The Software Defined Slicing (SDS) architecture introduced in~\cite{Gkelias2021} provides an enabling framework for managing DTs in IoBT coalitions through its innovative three-level control structure. This architecture facilitates flexible and dynamic resource allocation across coalition domains, crucial for effective DT deployment. It enables the formation of a coalition network by interconnecting resources (communication, storage, computation, databases, sensors, etc.) from different partners, achieved by reserving a Software Defined Slice (SDS) — a set of resources and Virtual Network Functions (VNFs) spanning multiple domains — for a specific coalition mission.

\begin{figure}[htbp]
\centerline{\includegraphics[width=0.37\textwidth]{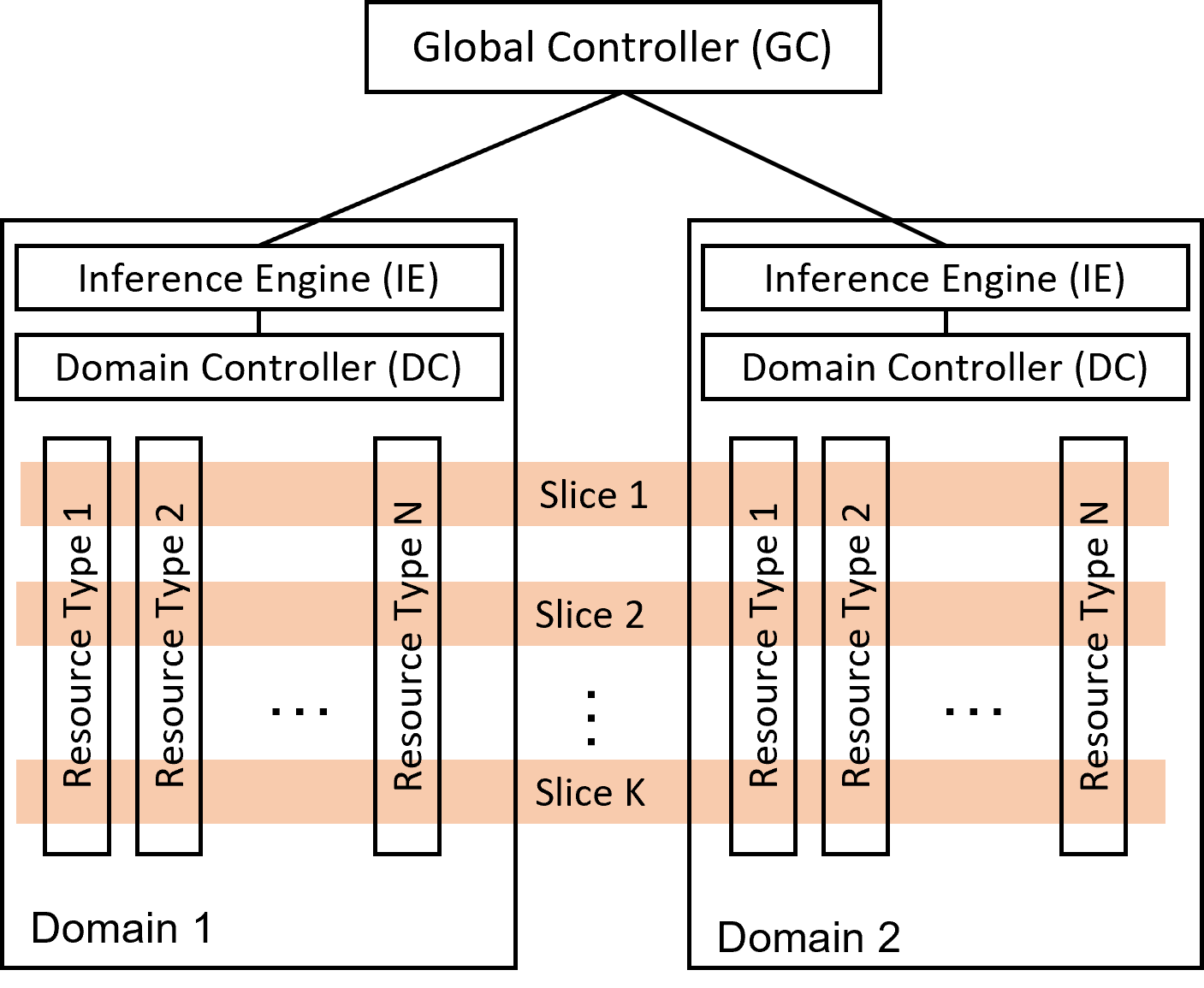}}
\caption{SDC architecture: GC, DC and IE interactions.}
\label{figSDS1}
\end{figure}

\begin{figure}[htbp]
\centerline{\includegraphics[width=0.37\textwidth]{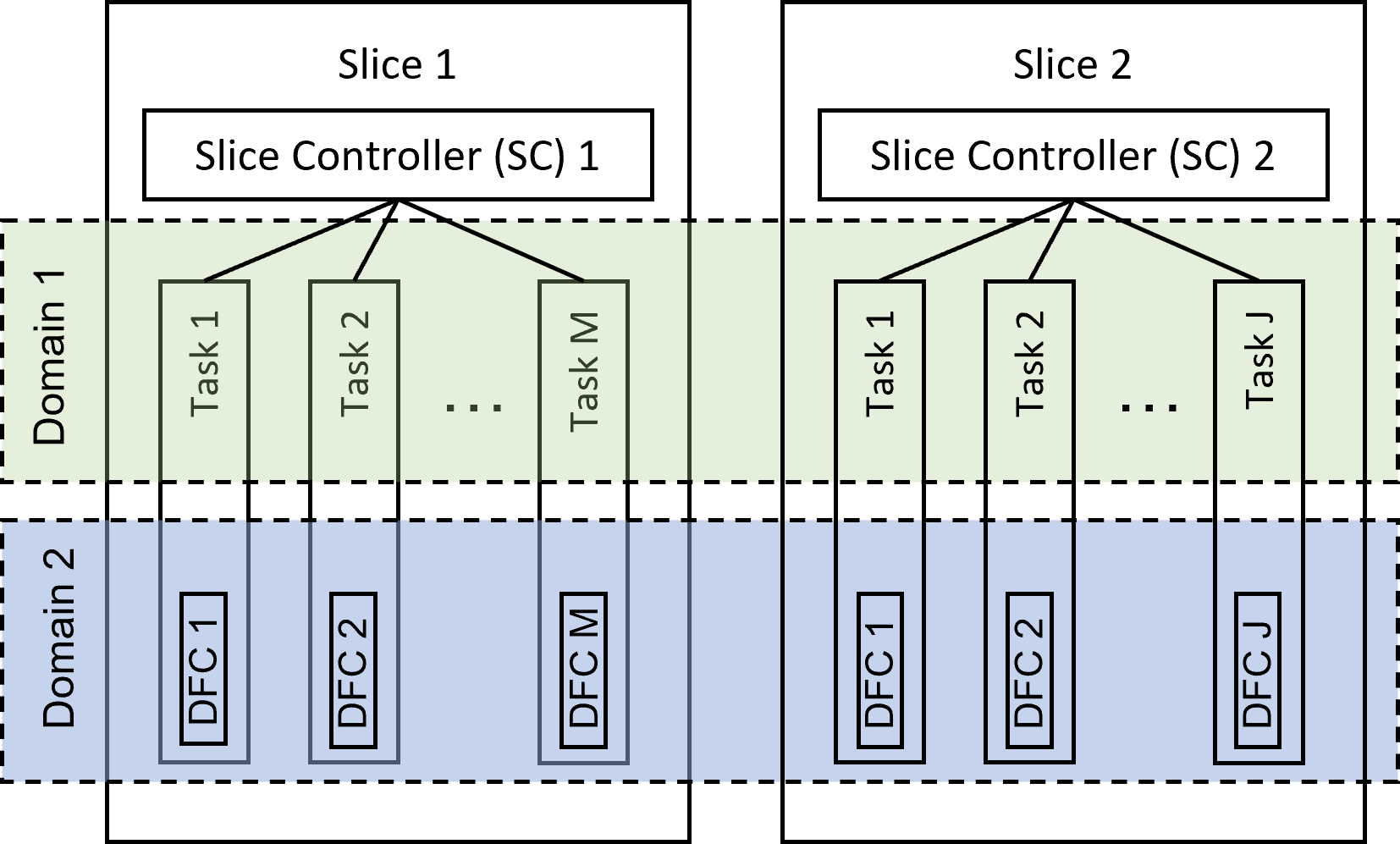}}
\caption{Slice Controller architecture.}
\label{figSDS2}
\end{figure}

A logically centralized Global Controller (GC) provide inter-domain negotiation and coordinates resource sharing among SDSs across these domains. The GC uses aggregated resource information from each domain, provided by Resource Inference Engines (IEs) associated with each Domain Controller (DC), to determine the feasibility of new SDS requests (Fig.~\ref{figSDS1}). If a request is accepted, the GC allocates resources across domains and instructs the DCs to reserve them. The GC operates without needing detailed knowledge of each domain's network topology, relying instead on the aggregated resource information reported by the IEs. The IEs learn and report the available capacity of aggregated resources without revealing specific domain details.
Each DC manages its domain's network topology and resource distribution. Upon receiving a resource reservation request from the GC, the DC decides whether it can fulfill the request. If accepted, the DC enforces the reservation within its domain. After all involved DCs accept their requests, a dedicated Slice Controller (SC) is created for the new SDS (Fig.~\ref{figSDS2}). The SC manages task acceptance and optimal resource allocation within the slice. Upon mission completion, the slice and SC are terminated. For each accepted task, a Dynamic Flow Controller (DFC) is activated to optimize data/processing rates and dynamically adjust to network changes and resource availability uncertainties.


The proposed DT architecture can integrate with the SDS framework to provide a holistic solution for IoBT coalitions, creating a mutually beneficial relationship: DT coalitions leverage SDS resource provisioning, while the SDS architecture benefits from DT-driven optimization, prediction, and real-time adaptation. Specifically, the GC and DTC-controller can collaborate to support DT model deployment based on computational and networking needs, enabling resource sharing between DTs and their physical counterparts (within the same slice) or dedicated resource allocation (separate slices) for sensitive models. DCs and IEs can leverage DTCP-controller capabilities by utilizing DT-generated synthetic data for AI training, incorporating accurate simulation models, and processing real-time IoBT inputs for optimal resource allocation and utilization. Furthermore, DTCM-controllers can collaborate with the corresponding SCs to optimize resource allocation within each slice, whether dedicated to DTs or shared with their physical counterparts. Finally, DFCs provide real-time resource allocation and adjustments, especially for edge-deployed DTs.

This combined SDS and DT framework provides a powerful tool for coalition forces, enabling flexible DT deployment and management across diverse battlespace environments while adhering to strict military security requirements. The architecture's adaptability, coupled with DT-enhanced decision support and resource optimization, delivers the scalability and resilience crucial for modern coalition warfare.

\section{Discussions and Future Directions}\label{sec:disc}

Developing efficient and robust DTs for IoBT coalitions presents several key research challenges. When standards and common data models exist, research on DT-controller design should focus on automatic discovery and registration of compliant DT models, including efficient validation methods and adaptive interfaces that accommodate evolving standards while maintaining backward compatibility. Implementing efficient, standardized data exchange protocols that optimize resource utilization is also crucial. In the absence of common standards, DT-controller design research should prioritize intelligent semantic mapping algorithms to automatically identify and resolve data format discrepancies between DT models. This includes developing dynamic data transformation techniques adaptable to varying data quality and formats from coalition partners, along with efficient caching and buffering for asynchronous data exchange between heterogeneous models. Robust error detection and correction for data transformation processes are also essential.

Synchronization and time management also represent a fundamental challenge requiring sophisticated solutions. Research should focus on developing algorithms for managing temporal consistency across distributed DT models and creating methods for handling varying update rates and latencies in coalition environments. This includes designing techniques for maintaining causal relationships between interdependent DT models and implementing efficient state synchronization mechanisms that minimize communication overhead while maintaining accuracy. 

The deployment of DT trained AI models to physical IoBT systems in coalition environments presents unique challenges due to the inherent discrepancies between simulated and real-world scenarios. While DTs provide valuable representations, they remain idealized models that may not fully capture real-world complexities, sensor inaccuracies, and environmental disturbances. This challenge is amplified in coalition environments where multiple defence partners share their DT IoBT models, as these models may vary in complexity, accuracy, and update frequency. The limited availability of real-world data in military environments, combined with the potential staleness of coalition DT outputs, can lead to significant performance degradation when deploying trained models to physical systems. To address these challenges, there is a pressing need for novel machine learning approaches that can effectively bridge the gap between DT-trained models and physical IoBT systems. These approaches must incorporate advanced transfer learning techniques that enable efficient model adaptation while considering the resource constraints of IoBT devices and the requirements for rapid deployment. Additionally, when sharing AI models between coalition partners with similar but non-identical physical systems, sophisticated transfer learning methods are needed to identify and transfer the most relevant features for model generalization while protecting sensitive information. These new approaches must balance communication and computational trade-offs, support online learning for continuous adaptation to changing environments, and maintain model effectiveness while ensuring generalization across coalition systems.

These research directions are essential for developing DT controllers that can effectively manage the complexities of coalition operations while ensuring interoperability, security, and performance in dynamic battlefield environments.

\section{Conclusions}\label{sec:con}

The integration of DTs within IoBT coalitions has become increasingly crucial as modern military operations demand enhanced collaboration, shared situational awareness, and efficient resource utilization across coalition partners. This paper has presented a comprehensive framework to address this need through a novel three-tier architecture that enables efficient coordination and management of DT models across coalition partners. The architecture leverages specialized controllers - DTCP controllers for managing individual partners' DT resources, a central DTC controller for cross-partner coordination, and DTCM controllers for mission-specific DT interactions - working in concert with software-defined networking principles to enable dynamic resource allocation and slice management. The proposed hybrid approach for DT model placement across edge devices, tactical nodes, and cloud infrastructure can optimise performance while maintaining security and accessibility. While this framework provides a robust foundation for deploying and managing DTs in coalition warfare, enhancing situational awareness and operational effectiveness, significant research challenges remain in areas such as interoperability, security, resource management, and AI model adaptation. These challenges, along with the rapid evolution of both DT and IoBT technologies, indicate that continued research and development efforts are essential to fully realize the potential of DT coalitions in future military operations.

\end{document}